\documentclass[aps,prb,
groupedaddress, showpacs]{revtex4}
\usepackage{amsmath}
\usepackage{amssymb}
\usepackage{graphicx}
\usepackage{color}
\definecolor{Blue}{rgb}{0.3,0.3,1}
\usepackage{tikz}
\usepackage{pgffor}
\usepackage{verbatim}
\begin{document}

\title{Fluctuation-dissipation theorem for chiral systems in non-equilibrium steady states}

\author{Chenjie~Wang and D.~E.~Feldman}
\affiliation{Department of Physics, Brown University, Providence, Rhode Island 02912, USA}

\date{\today}

\begin{abstract}

We consider a three-terminal system with a chiral edge channel
connecting the source and drain terminals. Charge can tunnel
between the chiral edge and a third terminal. The third terminal
is maintained at a different temperature and voltage than the
source and drain. We prove a general relation for the current
noises detected in the drain and third terminal. It has the same
structure as an equilibrium fluctuation-dissipation relation with
the nonlinear response $\partial I/\partial V$
in place of the linear conductance. The result applies to a
general chiral system and can be useful for detecting
``upstream'' modes on quantum Hall edges.
\end{abstract}


\maketitle

\section{Introduction}

Fluctuation-dissipation theorems (FDT) \cite{fdt} establish a beautiful and useful connection between response functions and noise.
They have a long history beginning with the Einstein relations and Nyquist formula and culminating in Kubo's linear response theory.
The standard FDT applies in thermal equilibrium only and much attention has been focused on its violations in nonequilibrium conditions. It became gradually clear that the FDT forms a special case of more general fluctuation theorems valid for various classes of nonequilibrium systems \cite{fdt}. Well-known examples are the Jarzynski equality \cite{J} and the Agarwal formula \cite{agarwal}.

The foundations of the linear response theory and the FDT are the Gibbs distribution and causality. According to the causality principle, there is a fundamental asymmetry between the past and the future since the future depends on the past but the past is not influenced by future events. This imposes crucial restrictions on response to any perturbations. In this paper we address chiral systems \cite{footnote1} which possess a similar asymmetry between left and right so that what happens on the right affects what later happens on the left but not {\it vice versa}. Obviously, this may only be possible if excitations can propagate just in one direction. Such chiral transport can occur in topological states of matter, a primary example being a low-temperature 2D electron gas in the conditions of the quantum Hall effect \cite{wen} (QHE). The gas is gapped in the bulk and its low-energy physics is determined by 1D chiral edge excitations. In the simplest QHE states all edge modes have the same chirality and hence the current flows in one direction only, e.g., clockwise. We show that in such chiral systems a Nyquist-type formula (\ref{1}) holds for the low-frequency current noise and {\it nonlinear} conductance even far from equilibrium. This far-from-equilibrium FDT is different from a more general Agarwal formula \cite{agarwal} for non-chiral systems which connects quantities that do not generally have an obvious physical meaning and cannot be easily extracted from experiment.

Our results apply beyond QHE. As usual in statistical mechanics, the simplest example of a chiral system comes from the physics of ideal gases. Consider a large reservoir filled with an ideal gas. A narrow tube with smooth walls and an open end is connected to the reservoir. Molecules can leave the reservoir through the tube. The projections of their velocities on the tube axis cannot change. Hence, they can only move from the reservoir to the open end of the tube and the system is chiral. Imagine now that molecules can escape through the walls of the tube with the probability depending on their position and velocity. A relation similar to Eq. (\ref{1}) can then be derived for the particle flux through the walls and the fluctuations of the fluxes through the walls and the open end of the tube. We discuss that relation in the appendix. The gas example is one-dimensional. Chiral systems are also possible in 2D. Indeed, topological states of matter with gapless chiral excitations on a 2D surface of a 3D system are possible (e. g., Ref. \onlinecite{FL} and related systems). In such systems, charge can propagate in both directions along one of the coordinate axes but only in one direction along the second axis. Moreover, chiral models can emerge beyond conventional condensed matter physics. For example, statistical mechanics has been used to describe traffic \cite{traffic}. A chiral model describes traffic on a network of one-way streets with no parking as long as no traffic jams form. 

Chiral edge states in QHE are of particular interest. It was proposed that non-Abelian anyons exist in QHE at some filling factors \cite{pf}, such as 5/2 and 7/2. If the prediction is true this will have major implications for fundamental physics and quantum information technology \cite{pf}. However, the nature of the 5/2 state remains an open question. Competing theories predict both Abelian and non-Abelian statistics [see Ref. \onlinecite{wb} for a review of proposed states].  Some of the proposed states have chiral edges and others do not. In particular, all published proposals for Abelian states are chiral \cite{footnote3}. Thus, testing chirality of QHE edges is important in this context \cite{feld-li,heiblum} and the theorem (\ref{1}) will be useful for that purpose. On the other hand, it is generally believed that the edges of the Laughlin states at $\nu=1/(2p+1)$ are chiral. This expectation is supported by the chiral Luttinger liquid model \cite{wen} (CLL). However, CLL faces major challenges from experiment (for a review, see Ref. \onlinecite{hr}). For example, it cannot explain observed quasiparticle transmission through an opaque barrier without bunching into electrons \cite{comforti,kf}. Thus, it is important to test major assumptions of CLL. One of them is chirality. Our theorem can be used for testing that assumption. Eq. (\ref{1}) has already been verified \cite{privite} in the limiting cases of $T\gg V$ and $V\gg T$.

\begin{figure}[h]
\includegraphics{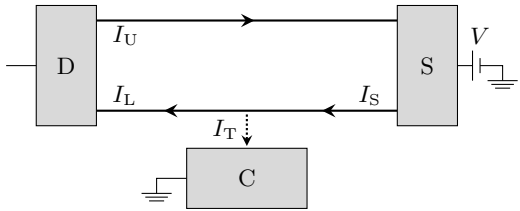}
\caption{Three-terminal setup. A quantum Hall bar is connected to source S at the voltage $V$. Charge tunnels into terminal C. The arrows represent the directions of the chiral edge modes.}
\end{figure}

A nonequilibrium FDT can be formulated for chiral systems in various geometries. Below we focus on the simplest geometry illustrated in Fig. 1. We consider a quantum Hall bar connected to the source (S) and drain (D) terminals. The impedance between the bar and the outside world is small. Long-range Coulomb forces are screened by a gate (this also ensures the absence of bulk currents \cite{bulk-curr}). Excitations propagate from the right to the left on the lower edge and in the opposite direction on the upper edge. The system size is much greater than the magnetic length; we assume that the chiral edges are far apart and do not influence each other. A third terminal C is connected to the lower edge through a tunneling contact. The details of the contact are unimportant and our results apply no matter how high or low the tunneling current $I_{\rm T}$
into terminal C is. A voltage bias $V$ is applied between the source and C. 
The temperature $T$ of the source and drain reservoirs may be different from the temperature of reservoir C. Our results do not depend on the latter temperature or the nature of conductor C. $V$ and $T$ should be much lower than the QHE gap (otherwise, QHE is absent and the system is not chiral).
We consider the current noise in terminal C,
$S_{\rm C}=\int dt \langle \Delta I_{\rm T}(t)\Delta I_{\rm T}(0)+\Delta I_{\rm T}(0)\Delta I_{\rm T}(t)\rangle$, and in the
 drain,
$S_{\rm D}=\int dt \langle \Delta I_{\rm D}(t)\Delta I_{\rm D}(0)+\Delta I_{\rm D}(0)\Delta I_{\rm D}(t)\rangle$, where  $I_{\rm D}$ is the electric  current in the drain and $\Delta I=I-\langle I\rangle$, and derive the relation
\begin{equation}
\label{1}
S_{\rm D}=S_{\rm C}-4T\frac{\partial I_{\rm T}}{\partial V}+4GT,
\end{equation}
where $G$ is the Hall conductance of the quantum Hall bar without a tunneling contact. This is the main result of the paper.

A similar formula in a different geometry was obtained in Refs. \onlinecite{a,b,c} for the exactly solvable CLL model. As seen below, Eq. (\ref{1}) holds independently of the integrability of a model and does not rely on CLL. Only chirality matters. This point is disguised in the model \cite{a,b,c} since the same solvable Hamiltonian describes a chiral system with tunneling between quantum Hall edges and a nonchiral quantum wire with an impurity.

We derive Eq. (\ref{1}) below. In the next section we give a simple heuristic derivation. Section III contains a full quantum proof of the fluctuation-dissipation theorem. We summarize our results and discuss their experimental implications in Section IV. Appendix contains a derivation of Eq. (\ref{1}) in an ideal gas model.

\section{Heuristic derivation}

Equation (\ref{1}) does not contain the Planck constant and so we first give its   heuristic classical derivation. The current $I_{\rm D}=I_{\rm L}-I_{\rm U}$ is composed of the current $I_{\rm L}$, entering the drain along the lower edge, and the current
$I_{\rm U}$ on the upper edge. These currents are uncorrelated and hence the noise in the drain is the sum of the noises of $I_{\rm U}$ and $I_{\rm L}$:
\begin{equation}
\label{2}
S_{\rm D}=S_{\rm U}+S_{\rm L}.
\end{equation}
The noise on the upper edge is the same as in the symmetric situation without tunneling into C. In the latter case the noise $S_{\rm D}$ is given by the Nyquist formula. Thus, $S_{\rm U}$ is one half of the equilibrium Nyquist noise,
$S_{\rm U}=2GT$.
In order to evaluate $S_{\rm L}$ we note that in a steady state there is no charge accumulation on the lower edge and hence the low-frequency component of the current, absorbed by the drain, $I_{\rm L}=I_{\rm S}-I_{\rm T}$, where $I_{\rm S}$
is the low-frequency part of the current, emitted from the source. Thus,
\begin{equation}
\label{3}
S_{\rm L}=S_{\rm C}+S_{\rm S}-2S_{\rm ST},
\end{equation}
where the cross-noise $S_{\rm ST}=\int dt \langle \Delta I_{\rm T}(t)\Delta I_{\rm S}(0)+\Delta I_{\rm S}(0)\Delta I_{\rm T}(t)\rangle$ and the noise of the emitted current equals one half of the Nyquist noise because of chirality,
\begin{equation}
\label{4}
S_{\rm S}=S_{\rm U}=2GT.
\end{equation}
We are left with the calculation of the cross-noise. The tunneling current depends on the average emitted current $GV$ and its fluctuations $I_{ \omega}$.
We assume that the central part of the lower edge has a relaxation time $\tau$. It is convenient to separate the fluctuations of $I_{\rm S}$ into fast, $I^{>}$, and slow, $I^{<}$, parts.
$I^{<}$ contains only frequencies below $1/\tau$. An instantaneous value of the tunneling current $I_{\rm T}(t)$
depends on the emitted current within the time interval $\tau$.  $I^{<}$ does not exhibit time-dependence within such time interval. Hence, it enters the expression for the tunneling current in the combination $GV+I^{<}$ only,
$I_{\rm T}=\langle I(GV+I^{<},I^{>})\rangle$, where the brackets denote the average with respect to the fluctuations of $I_{\rm S}$. According to the Nyquist formula for the emitted current, its  harmonics with different frequencies have zero correlation functions: $\langle I_\omega I_{-\omega'}\rangle\sim\delta(\omega-\omega')$.
 For the sake of the heuristic argument we will assume a Gaussian distribution of $I_S$ and hence independence of its high and low frequency fluctuations (no such assumptions are needed in a general proof).
After averaging with respect to the fast fluctuations of $I_{\rm S}$ we can write $I_{\rm T}=\langle J(GV+I^{<})\rangle$, where $J$ is obtained by averaging over $I^{>}$. $I^{<}$ corresponds to a narrow frequency window and can be neglected in comparison with $GV$, i.e., $I_{\rm T}=
J(GV)$. For the calculation of the cross-noise we expand $J(GV+I^{<})$ to the first order in $I^{<}$ and obtain
\begin{equation}
\label{5}
S_{\rm ST}=\langle I_{T,\omega}I_{-\omega}+I_{T,-\omega}I_{\omega}\rangle=\frac{\partial J(GV)}{G\partial V}\langle I_\omega I_{-\omega}+I_{-\omega}I_{\omega}\rangle=2T\frac{\partial I_{\rm T}}{\partial V},
\end{equation}
where we used the Nyquist formula for the fluctuations $I_{\omega}$ of $I_{\rm S}$. A combination of Eqs. (\ref{2}-\ref{5}) yields the desired result (\ref{1}) for nonequilibrium steady states.
Note that only low-frequency particle current conserves
along the edge in our screened conductor and hence it is plausible to expect that $I_{\rm T}$ depends only on harmonics $I_\omega$  with $\hbar\omega<\hbar\omega_{\rm cutoff}\ll T $.  In such case the assumption
of the Gaussian distribution and independence for $I_{<,>}$ is not needed since the same results can be obtained from the lowest order expansion in $I_{<,>}$.

The above argument easily generalizes to other geometries with many terminals and/or tunneling between QHE edges.

One can estimate noises $S_{\rm D}$ and $S_{\rm C}$. We assume that $V\sim T\approx 100mK$. These are realistic parameters for noise experiments. The injected current $I_S\sim e^2V/h$ since $G\sim e^2/h$. We assume that the transmission probability to conductor C is of the order of $1/2$. Thus, $I_T\sim I_S$. Then the second and third terms on the right hand side of Eq. (1) are of the order of $e^2[kT/h] \sim e^2[eV/h] $. The noise $S_C$ can be estimated from the fact that its physical meaning corresponds to the ratio of the fluctuation $\langle\Delta Q^2\rangle$ of the transmitted charge to
the time $\Delta t$ over which the charge was transmitted to C. The probability of a single tunneling event during the time interval $\Delta t \sim h/eV$ is of the order of $1/2$. Thus $S_C\sim e^2/\Delta t\sim e^2 [eV]/h$ and has the same order of magnitude as the second and third terms on the right hand side of Eq. (1). Finally, $S_D$ has the same order of magnitude. This corresponds to the noises of the order of $10^{-28}A^2/Hz$. This is a very small number but even lower noises are measured in the state of the art experiments in the field.

Counter-propagating edge modes break Eq. (\ref{1}). Imagine that an ``upstream" neutral modes propagates in the direction, opposite to that of the charged mode. Each tunneling event into C creates a neutral excitation
that brings energy $\sim V$ back to the source and heats it. This increases the noise, generated by the source, and raises the effective temperature above $T$ in Eq. (\ref{1}).  We would like to emphasize that the bulk of the source remains at the temperature $T$ since the heat capacity of the source is large. However, the bulk plays relatively little role in noise generation in our system due to a relatively low resistance of a massive bulk conductor.

\section{Proof of the nonequilibrium FDT}

We now turn to a full quantum derivation of Eq. (\ref{1}). To simplify notations we will omit the Boltzmann constant and $\hbar$ from the equations below.
As is clear from the above classical argument, the drain potential has no effect on the noises. It will be convenient to assume that the source and drain potentials are equal. It will also be convenient to assume that at the initial moment of time $t=-\infty$ there was no tunneling or other interactions between conductor C and the QHE subsystem containing the quantum Hall bar and the source and drain reservoirs. Thus, the system is initially described by the Hamiltonian $H_0=H_{\rm C}+H_{\rm H}$, where $H_{\rm C}$ and $H_{\rm H}$ denote the Hamiltonians of the two subsystems: conductor C and the QHE subsystem respectively. At later times the Hamiltonian includes an interaction term, $H=H_0+H_{\rm I}(t).$ 
One of the effects of the interaction is charge tunneling between the QHE subsystem and conductor C. 
We assume that the interaction $H_{\rm I}$ becomes time-independent well before the moment of time $t=0$ and the system is in its nonequilibrium steady state at $t=0$. The steady state depends on the voltage bias $V$, temperature $T$
and the temperature $T_{\rm C}$ of conductor C. All these energy scales must be lower than the QHE gap. Otherwise, the system is unlikely to allow a chiral description. We do not assume any special relation between the energy scales.
If $T=T_C$ and $V=0$ then the system is in equilibrium. If $T-T_C\ll T$ and $V\ll T$ then the system is close to equilibrium. Otherwise, the system is far from equilibrium. Our main result (\ref{1}) applies in all those cases.

\subsection{Chiral systems}

A general definition of a chiral system is the following: Consider a system     whose Hamiltonian has a time-dependent contribution
$H_t=\int_{-\infty}^ydxh(x,t)$, where the integration extends to the left of point $y$. In a chiral system, local observables to the right of  point $y$
do not depend on the form of  $h(x,t)$ for any initial conditions. 

In what follows, we will not need the most general definition. Instead, we will focus on one particular observable, current $I_{\rm S}$, emitted from the source to the lower edge. We are only interested in the low-frequency regime.
In that limit, the precise choice of the point, where the current $I_{\rm S}$ is measured, is unimportant due to the charge conservation. The same low-frequency current flows in all points of the lower edge between the source and conductor C. Let us select a point A in the gapped region in the bulk of the QHE liquid and a point B below the QHE bar (Fig. 2). The same current $I_{\rm S}$ must flow through any line, connecting A and B. This remains true, even if the line does not cross the lower edge and instead goes through the source (and the boundary between the source and the QHE liquid). Thus, $I_{\rm S}$ can be defined both in terms of the edge and source physics.

\begin{figure}[h]
\includegraphics{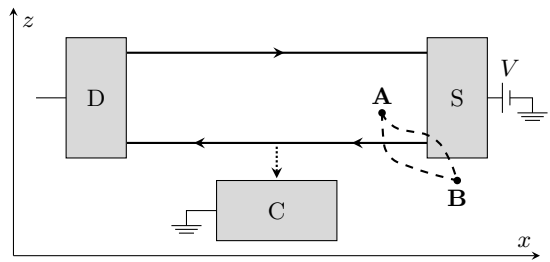}
\caption{The same low-frequency current $I_{\rm S}$ flows through both dashed lines.}
\end{figure}

The chirality assumption means that the average emitted current $I_{\rm S}(t)$ does not depend on the presence of conductor C for any initial conditions. In other words, the expression ${\rm Tr}\rho(t=-\infty)I_{\rm S}(t)$, where $\rho$ is the initial density matrix,
is the same when the Heisenberg operators 

\begin{equation}
\label{defISt}
I_{\rm S}(t)=[{\rm T}\exp(-i\int_{-\infty}^t V(t)dt)]^{-1}I_{\rm S}(-\infty)[{\rm T}\exp(-i\int_{-\infty}^t V(t)dt)],
\end{equation}
where $I_{\rm S}(-\infty)$ is a Schr\"odinger operator,
 are defined in terms of the Hamiltonians $V=H_0$ and 
$V(t)=H_0+H_{\rm I}(t)=H$. The latter can be true for a general $\rho(-\infty)$ only if the Heisenberg operators
$I_{\rm S}(t)$ are the same in the presence and absence of conductor C. Note that this property is satisfied in  the chiral Luttinger liquid model. To the right of conductor C, the CLL  action 
assumes the form $L=m\hbar/(4\pi)[\partial_t\phi\partial_x\phi-v(\partial_x\phi)^2]$,
where $v$ is the velocity of the edge excitations, the charge density $q=e\partial_x\phi/(2\pi)$ and the current operator $I_{\rm S}=vq$. From the equation of motion, $\partial_t\partial_x\phi-v\partial_x^2\phi=0$, we find that
the electric current on the right of conductor C, $I_{\rm S}=I_{\rm S}(t+x/v)$, depends only on the initial conditions on the right
and is not affected by the form of $H_{\rm I}(t)$. 
The same property is satisfied in any other chiral conformal theory and in many other situations.
For example, the chirality property of the operator $I_{\rm S}$ survives, if any changes are introduced into the above CLL action to the left of the point, where $I_{\rm S}$ is measured.
The chirality assumption also holds for QHE edges with several modes of the same chirality but breaks down, generally, if counter-propagating modes are present, as e.g., in the anti-Pfaffian state \cite{apf1,apf2} proposed at $\nu=5/2$. 

Edge reconstruction \cite{Wen-Chamon} may result in ``net chiral" edges that are not chiral. For example, a pair of counter-propagating integer QHE modes can emerge on a $\nu=1/3$ edge. In general, this breaks chirality.
However, in practice, disorder is likely to localize such mode pairs and restore chirality.

\subsection{Initial density matrix and Heisenberg current operator}

At the time $t=-\infty$ there is no interaction between conductor C and the QHE subsystem that includes the source, drain and 2D electron gas.
Hence, the initial density matrix $\rho(-\infty)=\rho_{\rm H}\rho_{\rm C}$ factorizes into a product of the initial density matrix $\rho_{\rm C}$ of conductor $C$ and the initial density matrix $\rho_{\rm H}$ of the QHE subsystem.
Each of them corresponds to an independent Gibbs distribution determined by an appropriate reservoir. At later times the subsystems interact, the steady state depends on both reservoirs, and the factorization property no longer holds
in the Schr\"odinger representation.
Thus, it will be convenient for us to perform calculations in terms of the initial density matrix because of its simpler structure \cite{foot3}. This means that we will use the Heisenberg formalism so that all time dependence is placed into 
the operators of observables. The chirality property will allow us to extract considerable information about the matrix elements of the Heisenberg operator $I_{\rm S}(t)$ and prove Eq. (\ref{1}). 
We would like to emphasize that the Hamiltonian has a time-dependent piece $H_{\rm I}(t)$ and this piece enters the definition of all Heisenberg operators. 
The presence of that piece is crucial for the difference between the density matrices in the Heisenberg and Schr\"odinger representations.
We will omit the time argument
in $\rho_{\rm H}$ and $\rho_{\rm C}$. It will be always understood that these are initial density matrices at $t=-\infty$. In all calculations below, $\rho$ is also taken at $t=-\infty$.
Certainly, in the Heisenberg representation, the density matrix does not depend on time.

Our approach resembles the Keldysh formalism, where all correlation functions are also expressed in terms of the initial density matrix. In the Keldysh technique, if the interaction is adiabatically turned on the initial density matrix describes free particles and hence factorizes into a product of single-particle density matrices. A difference from our approach consists in the application of the interaction representation in the Keldysh perturbation theory. The average of any properly time-ordered product of creation and annihilation operators is known exactly in the interaction representation. This allows development of a diagrammatic technique. We use the Heisenberg representation instead and rely on special properties of the matrix elements of the operator $I_{\rm S}$
in the basis, in which the initial density matrix is diagonal.

Any density matrix is Hermitian and can be diagonalized. Hence, 

\begin{equation}
\label{er1}
\rho_{\rm H,C}=\sum \rho_{{\rm H,C}n}|n_{\rm H,C}\rangle\langle n_{\rm H,C}|
\end{equation}
 and 

\begin{equation}
\label{er2}
\rho(-\infty)=\sum \rho_{n}|n\rangle\langle n |,
\end{equation}
 where

\begin{equation}
\label{er3}
\rho_n=\rho_{{\rm H}n'}\rho_{{\rm C}n''}
\end{equation}
and 

\begin{equation}
\label{er4}
|n\rangle=|n_{\rm H}'\rangle|n_{\rm C}''\rangle, 
\end{equation}
where the states $|n_{\rm H}'\rangle$ and $|n_{\rm C}''\rangle$ are selected from the Hilbert spaces of the QHE system and conductor C respectively.
$\rho_{\rm H}$ is a Gibbs distribution, $\rho_{{\rm H}n}\sim \exp(-E_n/T-|e|VN_n/T)$, where $N_n$ is the number of electrons in the QHE subsystem, $T$ the temperature of the source and drain reservoirs and $E_n$ are the eigenenergies of the eigenstates $|n_{\rm H}\rangle$ of the quantum Hall subsystem  {\it before} the tunneling contact was turned on, {\it i.e.}, $|n_{\rm H}\rangle$ are eigenstates of the Hamiltonian $H_{\rm H}$ with particle numbers $N_n$. We do not make assumptions about $\rho_{\rm C}$.
Our proof applies as long as the initial density matrix $\rho(-\infty)$ factorizes and the initial density matrix of the QHE subsystem $\rho_{\rm H}$ is given by the Gibbs distribution. In practice, the initial density matrix
$\rho_{\rm C}$ is also likely to be a Gibbs distribution. To avoid a possibility of confusion, we emphasize that all states in the bases $|n_{\rm H}\rangle$ and $|n_{\rm C}\rangle$ are time-independent. Thus, they are no longer eigenstates of the time-dependent Hamiltonian after
the interaction $H_{\rm I}$ has been turned on.

If the interaction $H_{\rm I}(t)$ is never turned on then $I_{\rm S}(t)$ acts in the Hilbert space of the QHE subsystem and hence its nonzero matrix elements are always diagonal in the basis of $|n_{\rm C}\rangle$.
The chirality property means that the same restriction applies to nonzero matrix elements of $I_{\rm S}(t)$ even {\it after} the interaction $H_{\rm I}(t)$ has been turned on since $I_{\rm S}(t)$ must be the same in the presence and absence of the interaction $H_{\rm I}(t)$.  The emitted current operator commutes with the number $N$ of the particles in the quantum Hall subsystem since it describes particle transfer between the source and the edge. Thus, in the absence of the interaction $H_{\rm I}(t)$, it has nonzero matrix elements only between states $|n_{\rm H}\rangle$  with the same $N_n$. Again, the chirality property means that the same restriction on nonzero matrix elements applies even {\it after} the interaction has been turned on.
  Before the interaction between the quantum Hall bar and subsystem C has been turned on, it is easy to write the time-dependence for matrix elements of any operator acting in the Hilbert space of the QHE subsystem: $\langle n|O_{\rm H}(t_1)|m\rangle=\exp(i[E_n-E_m](t_1-t_2))\langle n|O_{\rm H}(t_2)|m\rangle$. 
The same relation would apply at all times, if the interaction $H_{\rm I}(t)$ were never turned on. 
The chirality property means that the emitted current operator $I_{\rm S}(t)$ exhibits exactly the same time-dependence at any times, if the interaction $H_{\rm I}(t)$ is turned on and if it is not. This applies both before and 
{\it after} the tunneling between two subsystems has been turned on. Setting $t_2=0$ in the above relation, we obtain

\begin{equation}
\label{ISt}
\langle n|I_{\rm S}(t)|m\rangle=\exp(i[E_n-E_m]t)\langle n|I_{\rm S}(0)|m\rangle.
\end{equation}

\subsection{Voltage bias}

A standard way to include voltage bias in mesoscopic systems is based on the Landauer-B\"uttiker formalism: One assumes that the tunneling term is initially absent and then turned on and that the lower edge is initially at equilibrium with the reservoir with the chemical potential $V$. We will use a mixed Kubo-Landauer formalism to determine the response of $I_{\rm T}$ to a small change $\delta V$ of the voltage bias. It will allow us to reduce the problem of nonlinear response to $V$ to the linear response to $\delta V$.
In the mixed Kubo-Landauer language, an additional electromotive force $\delta V$ is generated by an infinitesimal time-dependent vector potential $\delta {\bf A}$. 

\begin{figure}[h]
\includegraphics{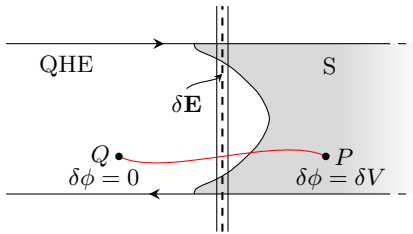}
\caption{(color online). Illustration of the bias voltage. $\delta{\bf A}$ and $\delta {\bf E}$ are applied in the region between two solid vertical lines. In the example in the figure the region with $\delta{\bf E}$ crosses both the source (shaded) and the gapped QHE region (white). $\delta\phi$ is constant on the vertical dashed line. $\delta\phi=0$ in point Q and $\delta\phi=\delta V$ in point P.}
\end{figure}

We assume that different contacts are connected with infinite reservoirs at different electrochemical potentials. Their difference determines the voltage bias $V$: the source electrochemical potential is $V$ and the potential of
conductor C is 0.
 A small change of the bias $\delta V$ can be introduced with an electric field described by a time-dependent vector potential, $\delta {\bf E}=-1/c d\delta {\bf A}/dt$. The electric field is applied in a finite part of the source terminal (Fig. 3) and cannot affect chemical potentials of the infinite reservoirs. The chemical potentials determine electric potentials of the reservoirs because of charge neutrality. The distribution of charges certainly changes in the middle of the conductor in the presence of $\delta {\bf A}$, so the electrostatic potential $\phi$ also changes. However, the potential difference between the reservoirs does not. The magnetic field must be time-independent inside the sample as required by the restrictions on e.m.f. sources in the circuit theory. In other words, $\delta{\bf B}= {\rm curl} \delta{\bf A}=0$ inside the conductor. 
Hence, the integral of $\delta {\bf E}$ does not depend on the choice of a path inside the conductor at fixed positions of its ends. If a path PQ begins in the infinite source reservoir and ends on the opposite side from the region with the field $\delta{\bf E}$ (Fig. 3) then $\int_P^Q d{\bf r}\delta{\bf E}=\delta V$. Hence, $\delta{\bf A}=ct\times{\rm grad}\delta\phi$, where $\delta\phi=\delta V$ in the source reservoir and $\delta\phi=0$ far on the left in the quantum Hall region (Fig. 3). As a consequence, the vector potential $\delta{\bf A}$ can be gauged out inside the conductor at the expense of changing the electrostatic potential $\phi\rightarrow\phi+\delta\phi$.  
This means a change of $\delta V$ in the electrochemical potential of the source reservoir and no change in the potential of conductor C. 
Thus, one can see that the Kubo formalism is equivalent to the Landauer-B\"uttiker approach in the presence of infinite electrically neutral reservoirs.

$\delta{\bf A}$ generates a correction to the Hamiltonian: $\delta H=-\int d^3 r \delta{\bf Aj}/c$, where ${\bf j}$ is the current density. Consider an arbitrary surface of constant $\delta\phi$ in the region with nonzero $\delta {\bf E}$ (Fig. 3). Similar to the discussion of Fig. 2, in the low-frequency limit, the total current through any such surface is the same and equals the total current through the source 

\begin{equation}
\label{totalcurrent}
I=I_{\rm U}-I_{\rm S},
\end{equation}
the signs in front of $I_{\rm U,S}$ reflecting our conventions about current directions, Fig. 1. This allows rewriting 

\begin{equation}
\label{deltaH}
\delta H=-I\delta \tilde A/c, 
\end{equation}
where
$d\delta \tilde A/dt=c\delta V$.
The same approach can be used to describe small changes of the drain potential but they are irrelevant for our purposes.

In what follows it will be convenient to consider the case of $\delta \tilde A$ oscillating with a low frequency $\omega$, $\delta \tilde A=c\delta V{\sin\omega t}/{\omega}$.

\subsection{Main argument}

We now give a full quantum derivation of Eq. (\ref{1}). The arguments leading to Eqs. (\ref{2}-\ref{4}) do not change compared to Section II and we concentrate on Eq. (\ref{5}). The cross-noise can be expressed as
\begin{align}
\label{6}
& 2S_{\rm ST}=\int dt \langle I_{\rm T}(0)I_{\rm S}(t)+h.c.\rangle(\exp(i\omega t)+\exp(-i\omega t)) = \nonumber\\
& \int dt \sum_{mn} (\exp(i\omega t)+\exp(-i\omega t))[\langle m|I_{\rm T}(0)|n\rangle\langle n|I_{\rm S}(t)|m\rangle\rho_m(-\infty)+\langle n|I_{\rm S}(t)|m\rangle\langle m|I_{\rm T}(0)|n\rangle\rho_n(-\infty)],
\end{align}
where a low frequency $\omega<1/\tau$ ($\tau$ is the relaxation time, Section II), $\rho(-\infty)$ is the initial density matrix and $I_{\rm T,S}$ are Heisenberg operators [see Eq. (\ref{defISt})]. 
As usual, introducing a small nonzero frequency allowed us to write the expression in terms of $I_{\rm S,T}$ and not $\Delta I_{\rm S,T}=I_{\rm S,T}-\langle I_{\rm S,T}\rangle$.
Eq. (\ref{6}) gives the noise at $t=0$, when the system is in a steady state.
As discussed in Section III.B it is convenient to use the Heisenberg representation in which the density matrix is not the steady state density  matrix $\rho(t=0)$ but the  initial $\rho(t=-\infty)$ since $I_{\rm S}(t)$  exhibits remarkable properties in such representation. Inserting the time dependence (\ref{ISt}) of the matrix elements $\langle n|I_{\rm S}(t)|m\rangle$, one finds
\begin{equation}
\label{6a}
2S_{\rm ST}=2\pi\sum_{mn}[\rho_m(-\infty)+\rho_n(-\infty)]\langle m|I_{\rm T}(0)|n\rangle\langle n|I_{\rm S}(0)|m\rangle[\delta(E_n-E_m+\omega)+\delta(E_m-E_n+\omega)].
\end{equation}

Next, we need to compute $R_{\rm T}=\partial I_{\rm T}/\partial V$. As discussed in Section III.C, this is a linear response problem with respect to $\delta V$.
Similar to Ref. \onlinecite{safi}, $R_{\rm T}$ is given by the same Kubo formula as in equilibrium. Indeed, 

\begin{equation}
\label{current1}
\langle I_{\rm T}(t=0)\rangle=
{\rm Tr}[\rho(-\infty)S_{\rm A}(-\infty,0)I_{\rm T}^s S_{\rm A}(0,-\infty)],
\end{equation}
 where $I_{\rm T}^s$ is a Schr\"odinger operator, $S_{\rm A}(t_2,t_1)$ the evolution operator, $S_{\rm A}(0,-\infty)={\rm T}\exp(-i\int^0_{-\infty}H_{A}(t)dt)$,
the Hamiltonian $H_A(t)=H-I\delta\tilde A/c$ and $H=H_{\rm C}+H_{\rm H}+H_{\rm I}(t)$. The expansion to the first order in $\delta\tilde A$ yields

\begin{equation}
\label{current2}
R_{\rm T}\times\delta V=
i\int_{-\infty}^0 dt{\rm Tr}[\rho(-\infty)\{S(-\infty,t)\delta H^s S(t,0)I_{\rm T}^sS(0,-\infty)-S(-\infty,0)I_{\rm T}^sS(0,t)\delta H^sS(t,-\infty)\}],
\end{equation}
where $\delta H^s$ is the Schr\"odinger operator (\ref{deltaH}) and $S(b,a)={\rm T}\exp(-i\int_{a}^b dt [H_{\rm C}+H_{\rm H}+H_{\rm I}(t)])$.
Substituting (\ref{deltaH}) in the above equation we see that the response of $I_{\rm T}$ to $\delta V$ expresses as the sum of the responses of $I_{\rm T}$ to the perturbations $I_{\rm S}\delta \tilde A/c$ and 
and $-I_{\rm U}\delta\tilde A/c$. The latter response is zero since the edges are far apart and perturbations on the upper edge have no effect on the lower edge.
With this in mind, we rewrite the
nonlinear response to $V$ in the form
\begin{equation}
\label{7}
R_{\rm T}=\partial I_{\rm T}/\partial V=i\lim_{\omega\rightarrow 0}\int^0_{-\infty} dt \sum_{mn} \frac{\exp(i\omega t)- \exp(-i\omega t)}{2i\omega}[\langle n|I_{\rm S}(t)|m\rangle\langle m|I_{\rm T}(0)|n\rangle\rho_n(-\infty)-\langle m|I_{\rm T}(0)|n\rangle\langle n|I_{\rm S}(t)|m\rangle\rho_m(-\infty)].
\end{equation}
In the above equation we absorbed evolution operators into the Heisenberg current operators.

It is convenient to combine the above response with the response $R_{\rm S}$ of $I_{\rm S}$ to the perturbation $\delta V I_{\rm T}\sin (\omega t)/\omega$ in the Hamiltonian. Certainly, that response is zero because of chirality.
Indeed, we consider a perturbation, acting on the left of the point, where $I_{\rm S}$ is measured.
 We get an expression of the same structure
as above with the indices $\rm S$ and $\rm T$ exchanged. In a steady state we expect that $\langle I_{\rm S}(0) I_{\rm T}(t)\rangle=\langle I_{\rm S}(-t) I_{\rm T}(0)\rangle$. This allows rewriting $R_{\rm S}$ in the form
\begin{equation}
\label{8}
R_{\rm S}=i\lim_{\omega\rightarrow 0}\int_0^{+\infty} dt \sum_{mn}\frac{\exp(i\omega t)- \exp(-i\omega t)}{2i\omega}[\langle n|I_{\rm S}(t)|m\rangle\langle m|I_{\rm T}(0)|n\rangle\rho_n(-\infty)-\langle m|I_{\rm T}(0)|n\rangle\langle n|I_{\rm S}(t)|m\rangle\rho_m(-\infty)].   
\end{equation}
We next compute $R_{\rm T}=R_{\rm T}+R_{\rm S}$:
\begin{equation}
\label{9}
R_{\rm T}=2\pi\lim_{\omega\rightarrow 0}\sum_{mn}\delta(E_n-E_m+\omega)\frac{\rho_n(-\infty)-\rho_m(-\infty)}{2\omega}[\langle n|I_{\rm T}(0)|m\rangle\langle m|I_{\rm S}(0)|n\rangle+\langle m|I_{\rm T}(0)|n\rangle\langle n|I_{\rm S}(0)|m\rangle],
\end{equation}
where we used the time-dependence (\ref{ISt}). The above equation contains the initial density matrix at time $t=-\infty$ and the Heisenberg current operators (\ref{defISt}) at time $t=0$.
Finally we apply the results of Section III.B for $I_{\rm S}(t)$ and $\rho(-\infty)$. We notice that nonzero matrix elements $\langle m|I_{\rm S}(0)|n\rangle$ correspond to $N_n=N_m$ and $|n_{\rm C}\rangle=|m_{\rm C}\rangle$.
Hence, in the limit of low frequencies in Eq. (\ref{9}), $[\rho_n(-\infty)-\rho_m(-\infty)]/\omega=-\rho_{{\rm C}n}d\rho_{{\rm H}n}/dE_n=\rho_n(-\infty)/T$, where we used the factorization property (\ref{er3})
which is only valid for the initial density matrix. Comparison of Eqs. (\ref{6a}) and (\ref{9}) at small $\omega$ establishes Eq. (\ref{5}).

The above calculation relies on the structure of the initial density matrix $\rho(-\infty)$. This does not mean that the steady state depends on minor details of the initial state. Only the temperatures and chemical potentials of the large reservoirs are important. Those temperatures and potentials remain the same in the initial and steady state. If, on the other hand, one of the reservoirs is not large then the steady state does not depend on the initial density matrix
of that reservoir.
This can be easily seen from Eq. (\ref{1}) in the limit of a small reservoir attached to conductor C. Indeed, in that case, $I_{\rm T}=S_{\rm C}=0$ in a steady state since conductor C cannot accumulate charge. Thus, Eq. (\ref{1})
reduces to $S_{\rm D}=4GT$. This is a usual Nyquist formula, valid for a system in thermal equilibrium at the temperature $T$ and a uniform chemical potential. Obviously, the steady state is indeed an equilibrium state with the temperature $T$, if conductor C is attached to a finite reservoir.  In this example, the final state does not depend on the initial density matrix $\rho_{\rm C}$. 

A similar argument does not work for a finite source reservoir. Indeed, the derivation of Eq. (\ref{1}) relies on the assumption that $I_{\rm U}$ is uncorrelated with $I_{\rm S}$. If the source reservoir is not large then the assumption is violated in the steady state and $I_{\rm U}=I_{\rm S}$ instead.

\section{Discussion}

The focus of the preceding section is on QHE, but similar non-equilibrium FDT apply in many other systems. The simplest example of a chiral system, based on an ideal gas, is considered in the appendix. Our results can also be generalized beyond 1D, for example, for the surface transport in a 3D stack of QHE systems. 

The geometry of Fig. 1 allows only electron tunneling to conductor C. FDT's, similar to (\ref{1}), can also be derived in other geometries, where fractionally charged anyons tunnel: One can consider tunneling between two edges of the same QHE liquid. 

Eq. (\ref{1}) does not contain the temperature $T_{C}$ of conductor C. This, certainly, does not mean that the properties of the system do not depend on it. The current $I_{\rm T}$ and the noises $S_{\rm D}$ and $S_{\rm C}$
are all affected by the temperature of C. The general relation (\ref{1}), however, remains the same. We would like to emphasize that our derivation does not contain any assumptions about the character of the dependence of $I_{\rm T}$
and $S_{\rm C}$ on the temperature and voltage. An interesting situation is possible, if conductor C is chiral and $T_C\ne T$. One can then derive two equations of the structure
(\ref{1}) with two different temperatures in them.

\begin{figure}[h]
\includegraphics{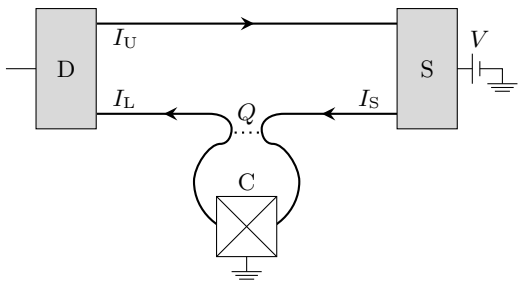}
\caption{A possible experimental setup. Charge carriers, emitted from the source, can either tunnel through the constriction Q and continue towards the drain or are absorbed by the Ohmic contact C.}
\end{figure}

Our main result, Eq. (\ref{1}), applies in chiral systems and can be used for an experimental test of chirality. A convenient measurement setup is illustrated in Fig. 4. Several mechanisms break chirality and can lead to the violation of Eq. (\ref{1}). One mechanism involves long range forces in the 2D electron gas. Our discussion assumed that a gate screens long-range Coulomb interaction. This allowed us to assume that the tunneling Hamiltonian
$H_{\rm I}$ does not depend on the voltage bias and the bias manifests itself in the 2D electron gas only through the chemical potential of the lower edge. Without screening, $H_{\rm I}$ may depend explicitly on the voltage and this must be taken into account at the calculation of $\partial I_{\rm T}/\partial V$. Strong interaction of edge modes with non-chiral bulk modes may also break chirality.

The most interesting mechanism of chirality breaking involves ``upstream" 
modes \cite{feld-li,heiblum}, Fig. 5. In the simplest example, two charged modes carry charge in the opposite directions. Let us imagine that the two chiral channels do not interact and all charge tunneling into C occurs due to particles, populating the upstream channel, directed from D to S. Then the noise in the drain is the same as in the absence of C, in contradiction with Eq. (\ref{1}).

\begin{figure}[h]
\includegraphics{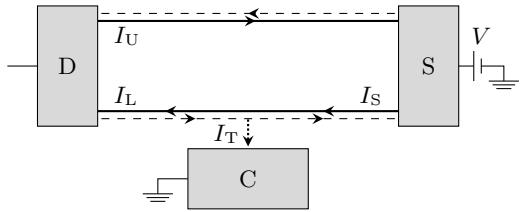}
\caption{A non-chiral system. The solid line along the lower edge illustrates the ``downstream mode", propagating from the source to the drain. The dashed line shows a counter-propagating ``upstream" mode.}
\end{figure}

This discussion neglects a possible heating effect. To illustrate it, let us assume that the upstream mode is neutral. It cannot carry charge but carries energy. In general, the tunneling operator into C includes a product of operators creating charged and neutral excitations on the edge. A neutral excitation of the energy $\sim V$ travels to the source and heats it. This affects noise, generated by the source, and leads to the violation of Eq. (\ref{1}). The details of the interaction of a neutral quasiparticle and the source are poorly understood theoretically. The experiment suggests that the heating effect will be strong \cite{heating}. Thus, large deviations from Eq. (\ref{1})
can be expected in the presence of ``upstream" neutral modes.

Our only assumption about $V$, $T$ and $T_{C}$ was that they are much lower than the QHE gap. Otherwise, a chiral description is unlikely to apply. If the system is chiral we make no assumptions about the relation between $V$ and $T$.
Nevertheless, our main focus was on the regime with $V\sim T$. Eq. (\ref{1}) greatly simplifies and becomes less interesting in the opposite limits $V\gg T$ and $T\gg V$. In the former case, let us set $T$ to zero. Then
Eq. (\ref{1}) reduces to $S_{\rm D}=S_{\rm C}$. This relation reflects noiseless character of the emitted current. In the opposite limit, let us assume that $V=0$ and $T=T_{\rm C}$. Then the equilibrium FDT applies. 
$\partial I_{\rm T}/\partial V$ is now linear response. Hence, $S_{\rm C}=4T\partial I_{\rm T}/\partial V$. Finally, Eq. (\ref{1}) reduces to $S_{\rm D}=4GT$. This simple relation reflects the fact that the lower edge is in thermal equilibrium on both sides of the contact with C.

In conclusion, we established a non-equilibrium FDT (\ref{1}) for chiral systems,  both close ($V\ll T$) and far ($V>T$) from equilibrium. The result does not apply to non-chiral conductors and can be helpful in the search of counter-propagating modes on
quantum Hall edges \cite{feld-li,heiblum}.

\acknowledgments

We acknowledge helpful discussions with
M. P. A. Fisher, M. Heiblum, S. Kehrein and A. W. W. Ludwig.  This work
was supported by NSF under Grant No. DMR-0544116 and
BSF under Grant No. 2006371.

\appendix
\section{Ideal gas model}

In this appendix we address a non-equilibrium FDT for an ideal gas system, briefly discussed in the introduction. 

We consider a large reservoir filled with an ideal gas of non-interacting molecules at the temperature $T$ and chemical potential $\mu$. Molecules can leave the reservoir through a narrow tube with smooth walls (Fig. 6). Collisions with the tube surface are elastic and do not change the velocity projection on the tube axis. Thus, molecules only move from the reservoir to the open end of the tube and the system is chiral. Imagine now that molecules can escape through a hole in the wall of the tube. We derive a relation similar to (\ref{1}):

\begin{equation}
\label{A1}
S_{\rm D}=S_{\rm C}-4T\frac{\partial I_{\rm T}}{\partial\mu}+S_{\rm S},
\end{equation}
where $I_{\rm T}$ is the particle current through the hole in the tube wall, and $S_{\rm D}=\int dt \langle \Delta I_{\rm D}(t)\Delta I_{\rm D}(0)+\Delta I_{\rm D}(0)\Delta I_{\rm D}(t)\rangle$,
$S_{\rm S}=\int dt \langle \Delta I_{\rm S}(t)\Delta I_{\rm S}(0)+\Delta I_{\rm S}(0)\Delta I_{\rm S}(t)\rangle$, and $S_{\rm C}=\int dt \langle \Delta I_{\rm C}(t)\Delta I_{\rm C}(0)+\Delta I_{\rm C}(0)\Delta I_{\rm C}(t)\rangle$  are respectively the particle current noises at the open end of the tube, at the opposite end of the tube, and at the hole (Fig. 6). The noise $S_{\rm S}$ can be determined from the measurement of $S_{\rm D}$ in the geometry without a hole in the tube wall. 

\begin{figure}[h]
\includegraphics{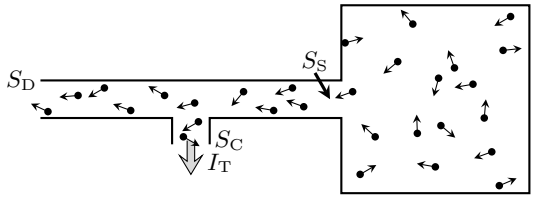}
\caption{Ideal gas in a reservoir with a tube.}
\end{figure}

The simplest proof of Eq. (\ref{A1}) is a direct calculation along the lines of Ref. \onlinecite{MartLan}. The calculation is especially simple in the case of an ideal classical gas which should be understood as a Fermi gas with a high negative chemical potential in order to use the above reference. Quantum Fermi- and Bose-gases are also easy to consider. The current and noise can be expressed as sums of contributions from 
different small energy intervals. Let $f=1/[\exp(\{E-\mu\}/T)+1]$ be the Fermi distribution function for a particular energy and $T_E$ the transmission coefficient through the tube wall for that energy ($T_E$ may depend on the channel number, if there are many channels). According to Ref. \onlinecite{MartLan}, in a Fermi gas, the contribution from a corresponding energy window to     the current, tunneling through the walls,  is proportional to $T_Ef$, the contribution to $S_{\rm S}$ is determined by $2f(1-f)$, $S_{\rm C}$ by $2T_Ef(1-T_Ef)$ and $S_{\rm D}$ by $2(1-T_E)f(1-(1-T_E)f)$. A combination of these contributions gives the desired theorem (\ref{A1}). 

One can also generalize our QHE proof. This approach, certainly, is harder than a direct calculation.
The situation simplifies for a degenerate Fermi gas whose particles can tunnel outside the tube only for energies, close to the Fermi level. 
Such gas can be mapped onto a model of charged particles whose mutual interaction is completely screened by the gate.
 The electric current and noise of such charged particles equal their mass current and noise up to a trivial coefficient. The connection of the chemical potential and voltage bias is obvious. 
Such model would describe left-moving electrons in a quantum wire in the language of the Landauer-B\"uttiker formalism. Its low-energy effective Hamiltonian is related to the integer QHE edge physics.
Tunneling through the tube walls plays exactly the same role as the tunneling into conductor C in the QHE setting.
The only important difference from a QHE setting, Fig. 1, is the absence of the upper edge. Thus, the derivation from the paper can be repeated with only one modification: $S_{\rm U}$ should be set to zero in Eq. (2). A small modification involves then Eq. (4): now $S_{\rm S}$ simply equals the noise at the open end of the tube in the absence of the hole in its side. That quantity must be substituted instead of $4GT$ in Eq. (1). Nothing else changes in that equation.


\begin{thebibliography}{100}
\bibitem{fdt} U. M. B. Marconi, A. Puglisi, L. Rondoni, and A. Vulpiani, Phys. Rep. {\bf 461}, 111 (2008).

\bibitem{J} C. Jarzynski, Phys. Rev. Lett. {\bf 78}, 2690 (1997).

\bibitem{agarwal} G. S. Agarwal, Z. Phys. {\bf     252}, 25 (1972).

\bibitem{footnote1} Consider a system whose Hamiltonian has a time-dependent contribution
$H_t=\int_{-\infty}^ydxh(x,t)$, where the integration extends to the left of point $y$. In a chiral system, local observables to the right of  point $y$
do not depend on the form of  $h(x,t)$ for any initial conditions.

\bibitem{wen} X.-G. Wen, {\it Quantum field theory of many-body systems} (Oxford, 2004).


\bibitem{FL} M. Levin and M. P. A. Fisher, Phys. Rev. B {\bf 79}, 235315 (2009).

\bibitem{traffic} D. Helbing, Rev. Mod. Phys. {\bf 73}, 1067 (2001).

\bibitem{pf} C. Nayak, S. H. Simon, A. Stern, M. Freedman, and S. Das Sarma, Rev. Mod. Phys. {\bf 80}, 1083 (2008).

\bibitem{wb}  B.J. Overbosch and X.-G. Wen, arXiv:0804.2087 (unpublished).

\bibitem{footnote3} Pairs of counter-propagating modes, which may emerge from edge reconstruction, are likely to be localized by disorder.


\bibitem{feld-li} D. E. Feldman and F. Li, Phys. Rev. B {\bf 78}, 161304(R) (2008). 

\bibitem{heiblum} A. Bid, N. Ofek, H. Inoue, M. Heiblum, C. L. Kane, V. Umansky, and D. Mahalu, Nature {\bf 466}, 585 (2010).

\bibitem{hr} M. Heiblum, arXiv:0912.4868 (unpublished).

\bibitem{comforti} E. Comforti, Y.C. Chung, M. Heiblum, V. Umansky, and D. Mahalu,
Nature (London) {\bf 416}, 515 (2002).

\bibitem{kf} C. L. Kane and M. P. A. Fisher, Phys. Rev. B {\bf 67}, 045307 (2003).

\bibitem{privite} M. Heiblum, private communication.

\bibitem{bulk-curr} C. L. Kane and Matthew P. A. Fisher,
Phys. Rev. B {\bf 52}, 17393 (1995). 


\bibitem{a} C. L. Kane and M. P. A. Fisher, Phys. Rev. Lett. {\bf 72}, 724 (1994).

\bibitem{b} P. Fendley, A. W. W. Ludwig, and H. Saleur, Phys. Rev. Lett. {\bf 75}, 2196 (1995).

\bibitem{c} P. Fendley and H. Saleur, Phys. Rev. B {\bf 54}, 10845 (1996).

\bibitem{apf1} M. Levin, B. I. Halperin, and B. Rosenow, Phys. Rev. Lett. {\bf 99}, 236806 (2007).

\bibitem{apf2} S.-S. Lee, S. Ryu, C. Nayak, and M. P. A. Fisher, Phys. Rev. Lett. {\bf 99}, 236807 (2007). 

\bibitem{Wen-Chamon} C. de C. Chamon and X. G. Wen,
Phys. Rev. B {\bf 49}, 8227 (1994). 

\bibitem{foot3} Strictly speaking, the assumption of factorization at $t=-\infty$ is not necessary. It is sufficient to assume that the final steady state depends only on the states of the reservoirs and not on the initial state of the finite central part of the system. In such case it is most convenient to perform calculations for the initial state whose density matrix factorizes. Certainly, the latter assumption itself is completely standard and can be easily tested experimentally by comparing steady states prepared from different initial conditions at the same temperatures and chemical potentials of the reservoirs.

\bibitem{safi} I. Safi, arXiv:0908.4382 (unpublished).

\bibitem{heating}
Y. Gross, M. Dolev, M. Heiblum, V. Umansky, and D. Mahalu, arXiv:1109.0102 (unpublished).

\bibitem{MartLan} Th. Martin and R. Landauer, Phys. Rev. B {\bf 45}, 1742 (1992).



\end{thebibliography}
\end{document}